\documentclass[twocolumn,prc,showpacs,preprintnumbers,nofootinbib,%
superscriptaddress]{revtex4}
\usepackage{mathrsfs}
\usepackage{amssymb}
\usepackage{amsmath}
\usepackage{graphicx}
\usepackage{mciteplus}
\usepackage{hyperref}

\newcommand{\rt}{{\mathbf{r}_T}}
\newcommand{\xt}{{\mathbf{x}_T}}
\newcommand{\bt}{{\mathbf{b}_T}}
\newcommand{\yt}{{\mathbf{y}_T}}
\newcommand{\zt}{{\mathbf{z}_T}}
\newcommand{\pt}{{\mathbf{p}_T}}

\newcommand{\kt}{{\mathbf{k}_T}}

\newcommand{\ptt}{p_T} 
\newcommand{\ktt}{k_T} 

\newcommand{\ud}{\, \mathrm{d}}

\newcommand{\tr}{\, \mathrm{Tr} \, }

\newcommand{\nc}{{N_\mathrm{c}}}
\newcommand{\nf}{{N_\mathrm{F}}}

\newcommand{\cf}{C_\mathrm{F}}
\newcommand{\ca}{C_\mathrm{A}}

\newcommand{\nr}[1]{(\ref{#1})}

\newcommand{\gev}{\ \textrm{GeV}}
\newcommand{\fm}{\ \textrm{fm}}

\newcommand{\qs}{Q_\mathrm{s}}

\newcommand{\qsadj}{\widetilde{Q}_{\mathrm{s}}}
\newcommand{\qsoadj}{\widetilde{Q}_{\mathrm{s0}}}
\newcommand{\qso}{Q_\mathrm{s0}}

\newcommand{\lqcd}{\Lambda_{\mathrm{QCD}}}
\newcommand{\as}{\alpha_{\mathrm{s}}}

\newcommand{\subA}{A}
\newcommand{\subB}{B}
\newcommand{\A}{A_{\subA}}
\newcommand{\B}{A_{\subB}}

\newcommand{\fig}{Fig.~}
\newcommand{\figs}{Figs.~}
\newcommand{\eq}{Eq.~}

\begin{document}

\author{T. Lappi}
\affiliation{
Department of Physics, %
 P.O. Box 35, 40014 University of Jyv\"askyl\"a, Finland}
\affiliation{
Helsinki Institute of Physics, P.O. Box 64, 00014 University of Helsinki,
Finland}

\title{
Gluon spectrum in the glasma from JIMWLK evolution
}

\pacs{24.85.+p,25.75.-q,12.38.Mh}

\begin{abstract}
The JIMWLK equation with a ``daughter dipole'' running coupling is solved numerically,
starting from an initial condition given by the McLerran-Venugopalan model.
The resulting Wilson line configurations are then used to compute the spectrum of gluons 
comprising the glasma inital state of a high energy heavy ion collision. 
The development of a geometrical scaling region makes the spectrum 
of produced gluons harder. Thus the ratio of the  mean gluon 
transverse momentum to the saturation scale grows with energy.
Also  the total gluon multiplicity increases with energy slightly faster than the 
saturation scale squared.
\end{abstract}

\maketitle

\section{Introduction}

Particle production in the initial stage of a  high energy heavy collision,
at RHIC or the LHC, is dominated by the small $x$ gluonic degrees
of freedom in the nuclear wavefunctions. 
When the dynamics of these gluons is dominated by a semihard 
\emph{saturation scale} $\qs$ it is possible to understand
the initial particle production in terms of weak coupling,
first principles QCD. This can be done 
in the framework of the CGC effective theory (for reviews see
e.g.~\cite{Iancu:2003xm,*Weigert:2005us,*Gelis:2010nm,*Lappi:2010ek}),
where the calculation is organized in terms of the classical gluon field 
for the small $x$ degrees of freedom, radiated from an effective
color source representing the larger $x$ partons. This generic division
only relies on the large energy, which, due to time dilation,
makes the large $x$ degrees of freedom evolve slowly,
and on gluon saturation and weak coupling, which guarantees
the validity of the classical field approximation.

The energy (or $x$) dependence of the gluonic degrees of freedom is
encoded in the 
JIMWLK~\cite{Jalilian-Marian:1997xn,*Jalilian-Marian:1997jx,*Jalilian-Marian:1997gr,%
*Jalilian-Marian:1997dw,*JalilianMarian:1998cb,*Iancu:2000hn,%
*Iancu:2001md,*Ferreiro:2001qy,*Iancu:2001ad,*Mueller:2001uk}
renormalization group equation, which  is obtained by successively integrating 
out the quantum fluctuations around the classical field into the color source.
The small $x$ gluonic degrees of freedom are most conveniently described in terms
of Wilson lines in the classical field. The correlators of these Wilson lines
can be probed experimentally by scattering a dilute probe off the CGC, such as in 
DIS or pA collisions at forward rapidity. 
Many phenomenological applications of this framework use 
the BK~\cite{Balitsky:1995ub,*Kovchegov:1999yj} equation, which can be 
derived from  JIMWLK  in a mean field approximation.

A collision of two objects described in the CGC framework can also be understood
in terms of classical gluon fields, the glasma fields~\cite{Lappi:2006fp}.
They start off for $\tau \lesssim 1/\qs$ as longitudinal chromoelectric and 
magnetic fields and evolve for $\tau \gtrsim 1/\qs$ into modes
that can be described as gluons with transverse momenta of order $\qs$.
The basic structure of the glasma fields has been known for some 
time~\cite{Kovner:1995ts}.
They can be obtained from a solution of the 
 Classical Yang-Mills (CYM) equations of motion
starting  from an initial condition
expressed in terms of the Wilson lines mentioned above.
The full equations have not been solved analytically, but
there has been a significant amount of work 
to develop numerical solutions.
So far, however, the numerical 
CYM~\cite{Krasnitz:1998ns,*Krasnitz:2001qu,*Krasnitz:2003jw,Lappi:2003bi}
calculations have been performed using the Wilson lines from the 
MV model~\cite{McLerran:1994ni,*McLerran:1994ka,*McLerran:1994vd},
in stead of the full solution of the JIMWLK equation. 
The only calculations of gluon production to include the 
dynamics of high energy evolution 
have been done using a solution of the BK equation in a 
$\ktt$-factorized 
approximation~\cite{Gelis:2006tb,Albacete:2007sm,Dusling:2009ni,Albacete:2010bs,*Albacete:2010ad}.
Since  $\ktt$-factorization does not correctly give the initial gluon spectrum, 
or multiplicity, for the collision of two dense systems 
(the ``AA'' case)~\cite{Blaizot:2008yb,Blaizot:2010kh,Levin:2010zs}, these
calculations are insufficient to give a full picture of the initial 
gluonic matter produced in a heavy ion collision.

The goal of this paper is to take the missing step of combining JIMWLK 
evolution with a full CYM calculation of gluon production in a heavy ion collision.
The JIMWLK equation is solved numerically,  using a (``daughter dipole'')
running coupling constant. The resulting
Wilson line configurations are then  used as initial conditions in a numerical 
computation of the spectrum of gluons produced in a collision of 
two sheets of CGC. The point of view taken in this paper is that the MV model
should be a reasonable initial condition for the evolution around RHIC energies.
We shall show explicitly how, starting from this initial condition,
the effect of JIMWLK evolution is the creation of a geometrical scaling region 
for momenta $\ktt \gtrsim \qs$, with a gluon spectrum that is harder 
than in the initial condition. This feature is carried over 
from  the wavefunction to the spectrum of gluons in the glasma, making
the glasma initial state more energetic at higher energies 
than a straightforward extrapolation of the MV model.

The numerical method for solving the JIMWLK equation is the one developed in 
\cite{Rummukainen:2003ns} and the one for solving the CYM equations of motion that used
e.g. in Refs~\cite{Krasnitz:1998ns,*Krasnitz:2001qu,*Krasnitz:2003jw,Lappi:2003bi}.
We will thus only briefly describe them in Sec.~\ref{sec:method}, and refer the 
reader e.g. to Refs.~\cite{Rummukainen:2003ns,Lappi:2009xa,Blaizot:2010kh}
for a more extensive discussion. We will then characterize our results for 
the JIMWLK equation in Sec.~\ref{sec:resjimwlk} and for the CYM calculation 
in Sec.~\ref{sec:rescym}, before discussing phenomenological context
and future directions in Sec.~\ref{sec:disc}.

The BK/JIMWLK evolution is most conventionally analyzed in terms of the fundamental 
representation saturation scale, which we denote by $\qs$. Gluon production, 
on the other hand, is expected to be dominated  by the adjoint representation
saturation scale which we denote $\qsadj$. These are related by a simple
color factor $\qsadj^2 = [\ca/\cf] \qs^2 = [2\nc^2/(\nc^2-1)] \qs^2$. 
The exact definition of $\qs$
used here is given in terms of the coordinate space Wilson line correlator in
\eq\nr{eq:defqs}.

\section{Solving JIMWLK and the CYM equations of motion}
\label{sec:method}

The JIMWLK equation describes the rapidity (or energy) evolution
of the probability distribution for Wilson lines.
In the CGC framework large $x$ degrees of freedom are described by 
static color charges, which serve as sources for a classical color field.
Most physical observables can be expressed in terms of Wilson lines formed
from the color field (in the covariant gauge, for a source moving in 
the positive $z$ direction)
\begin{equation}
U(\xt) = P \exp \left\{ i \int \ud x^- A^+_{\mathrm{cov}}(\xt,x^-) \right\}.
\end{equation}
These  Wilson lines are random SU(3) matrices from a probability distribution 
$W_y[U(\xt)]$, which depends on the rapidity cutoff $y$ that separates
the large $x$ color sources from the small $x$ classical field.
 As the energy is increased, successive layers of 
quantum fluctuations  have to be integrated into the probability distribution.
This leads to the JIMWLK renormalization group equation
\begin{multline}
\partial_y W_y[U]= 
-
\frac{1}{2} \frac{\as}{\pi^2}
\int\limits_{\xt \yt \zt}
\frac{\delta}{\delta A_{\mathrm{cov}}^{c+}(\xt)}
\\
\bigg[ \left(1-U^\dag(\xt) U(\zt)\right)^{ca}
\left(1-U^\dag(\yt) U(\zt)\right)^{ba}
\\
\frac{(\xt-\zt)\cdot(\yt-\zt) }{(\xt-\zt)^2(\yt-\zt)^2}
\frac{\delta}{\delta A_{\mathrm{cov}}^{b+}(\yt)}
W_y[U]
\bigg]
,
\end{multline}
which describes the energy dependence of this probability distribution.
This evolution equation can be written in a Langevin form for the
rapidity evolution of the Wilson lines
\begin{equation}
U_{y + \ud y}(\xt) = U_y(\xt)e^{i \alpha^a(\xt,y) t^a},
\end{equation}
where at each timestep the Wilson line is rotated in color space by
\begin{eqnarray}
&& \alpha^a(\xt,y) = 
-
\int\limits_{\zt} 
\bigg[
\frac{i \as \ud y }{2 \pi^2 (\xt-\zt)^2} 
\tr\left[ T^a \tilde{U}^\dag(\xt) \tilde{U} (\zt) \right]
\nonumber \\ 
&& \,\, +
\frac{\sqrt{ \as \ud y}}{\pi}
\frac{(x-z)^i}{(\xt-\zt)^2}
\left[1-U^\dag(\xt) U(\zt)\right]^{ab}
\eta_i^b(\zt)
\bigg],
\label{eq:ustep}
\end{eqnarray}
with a random noise
\begin{equation}
\langle \eta_i^a(\xt,y) \eta_j^b(\yt,y') \rangle 
= 
\delta^{ij}\delta^{ab}\delta^2(\xt-\yt) \delta(y-y').
\end{equation}
Here $\tilde{U}$ denotes the matrix in the adjoint representation,
and $T^a$ are the adjoint representation generators.
This numerical procedure relies on the factorization of the JIMWLK 
kernel into a product of two terms, depending only on the coordinate
pairs $\xt,\zt$ and $\yt,\zt$ (the ``daughter dipoles''). 

Published numerical solutions of the JIMWLK 
equation so far~\cite{Rummukainen:2003ns,Kovchegov:2008mk}
have used a fixed coupling constant $\as$. This gives an evolution speed
(increase of $\qs$ with energy) that is too fast to be phenomenologically 
reasonable, so it is essential to use a running coupling constant here.
There has been much discussion on the best running coupling 
prescription for BK/JIMWLK evolution in the 
literature~\cite{Kovchegov:2006vj,Balitsky:2006wa,Albacete:2007yr}.
The running coupling constant resums a subset of the NLO corrections
to BK/JIMWLK evolution, and different prescriptions correspond to 
resumming a different subset of them. Thus 
there is no unique ``correct'' way to set the scale of the coupling constant,
although it has been argued~\cite{Albacete:2007yr} that the Balitsky
prescription~\cite{Balitsky:2006wa} minimizes the effect of other NLO 
corrections. 
The numerical Langevin method of solving the JIMWLK equation 
relies on factorizing the JIMWLK kernel into a product of two factors 
that only depend on the sizes of the ``daughter'' dipole $\xt-\zt$
in \eq\nr{eq:ustep}. Thus implementing the 
preferred prescription of \cite{Balitsky:2006wa} would be difficult 
in a numerical solution of JIMWLK. We therefore use in this work the ad hoc
``daughter dipole'' prescription, where the magnitude of 
the coupling in \eq\nr{eq:ustep} only depends on $\xt-\zt$.
One also has to regulate the Landau pole in the coupling constant, which 
we do in a smooth way at a scale $\mu_0$ by taking
\begin{equation}
\as(r) = \frac{12 \pi}{ \left(33-2\nf \right) 
\ln \left[ \left(\mu_0^2/\Lambda^2\right)^{1/c} 
+ \left(r^2 \Lambda^2/4\right)^{-1/c}\right]^c}
\end{equation}
with $\nf=3$ and $c=0.2$. The value of the frozen coupling is
$\alpha_0= \frac{12 \pi}{ (33-2\nf) \ln (\mu_0^2/\Lambda) }.$
The scale $\Lambda$ in the coupling is parametrically
of the order of $\lqcd$, but the exact value that should be used
is scheme dependent. In the running coupling BK fit to DIS 
data~\cite{Albacete:2009fh,*Albacete:2010sy} the scale is
taken as a fit parameter and the data is found to 
prefer a smaller value $\Lambda^2 = \lqcd^2/6.5,$
which we will assume here.

The initial Wilson lines at $y=0$ are taken from the MV model, 
using the method discussed in more detail e.g. in Ref.~\cite{Lappi:2007ku}.
After a given number of iterations of the Langevin 
equation~\nr{eq:ustep} (we use a step size $\ud y = 0.0001 \pi^2/\alpha_0$)
one then obtains the configurations at a higher rapidity $y$. This
is done separately for two independent configurations, corresponding
to the two colliding nuclei.
In this work we only consider the symmetric situation of 
particle production at midrapidity, thus we evolve
both nuclei starting from the same $\qso$ and for the same interval
in $y$. 

Denoting the two nuclei as $A$ and $B$ one proceeds by
constructing the light cone gauge fields 
corresponding to the Wilson lines as
\begin{equation}\label{eq:pureg}
A_i^{\subA,\subB} = \frac{i}{g}U^\dag_{\subA,\subB} \partial_i U_{\subA,\subB}.
\end{equation}
The CYM equations of motion for the glasma fields 
are then solved as an initial value problem with initial conditions~\cite{Kovner:1995ts}
\begin{eqnarray}\label{eq:initcond}
\left. A^i \right|_{\tau=0^+} &=& \A^i + \B^i, \\
\left. A^\eta \right|_{\tau=0^+} &=& \frac{ig}{2}[\A^i,\B^i].
\end{eqnarray}
At a given proper time of ($\tau=12/\qsadj$ in our case) these fields are 
then Fourier-decomposed into $\kt$-modes to determine the spectrum
of the produced gluons.

\section{Results for JIMWLK}
\label{sec:resjimwlk}

\begin{figure}[tb]
\begin{center}
\includegraphics[width=0.4\textwidth]{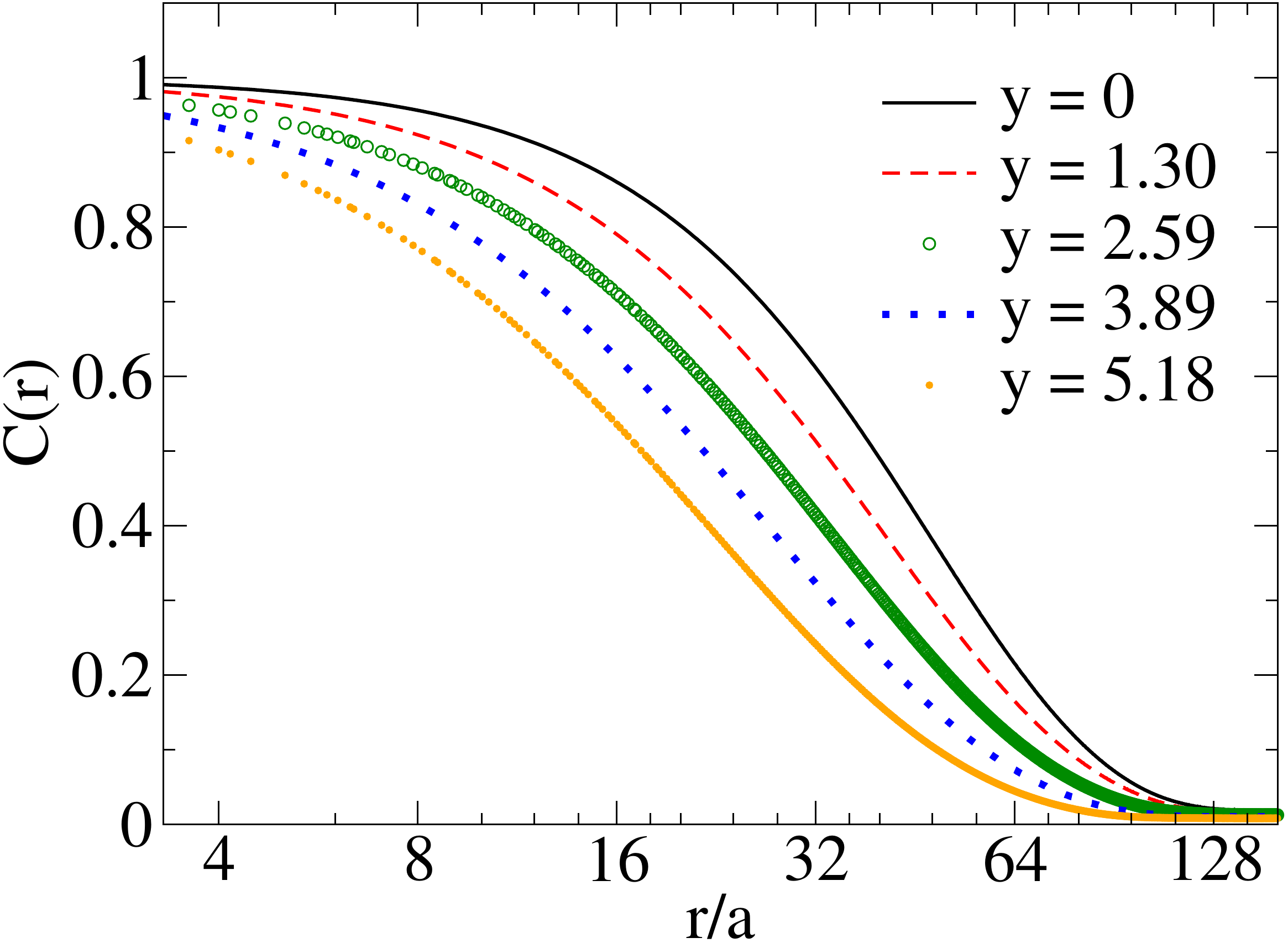}
\includegraphics[width=0.4\textwidth]{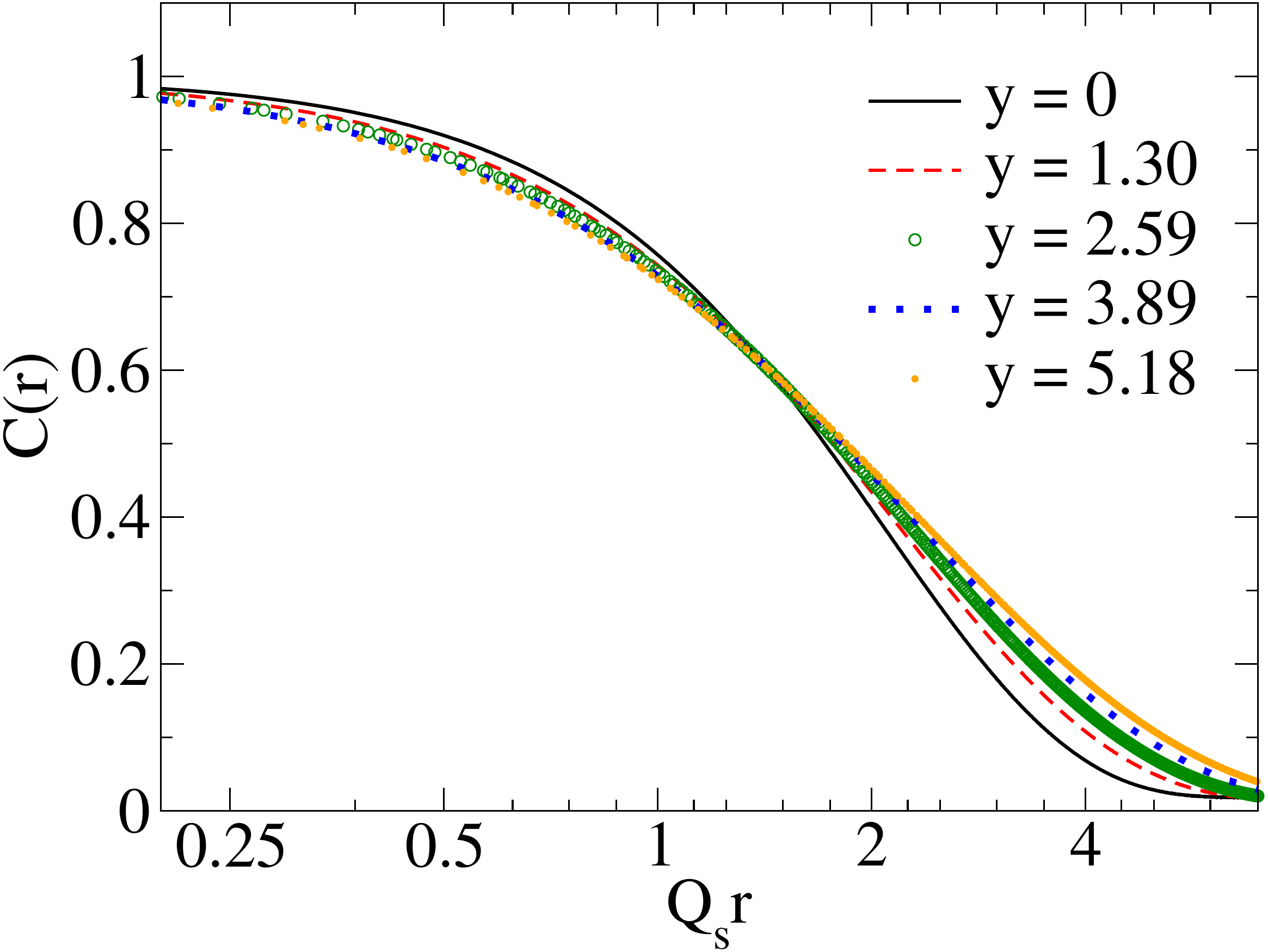}
\end{center}
\caption{
Wilson line correlator \nr{eq:defcorr}  in coordinate space; in lattice 
units (above) and as a function of the scaling variable $r \qs$ (below).
} \label{fig:uupos}
\end{figure}

\begin{figure}[tb]
\begin{center}
\includegraphics[width=0.4\textwidth]{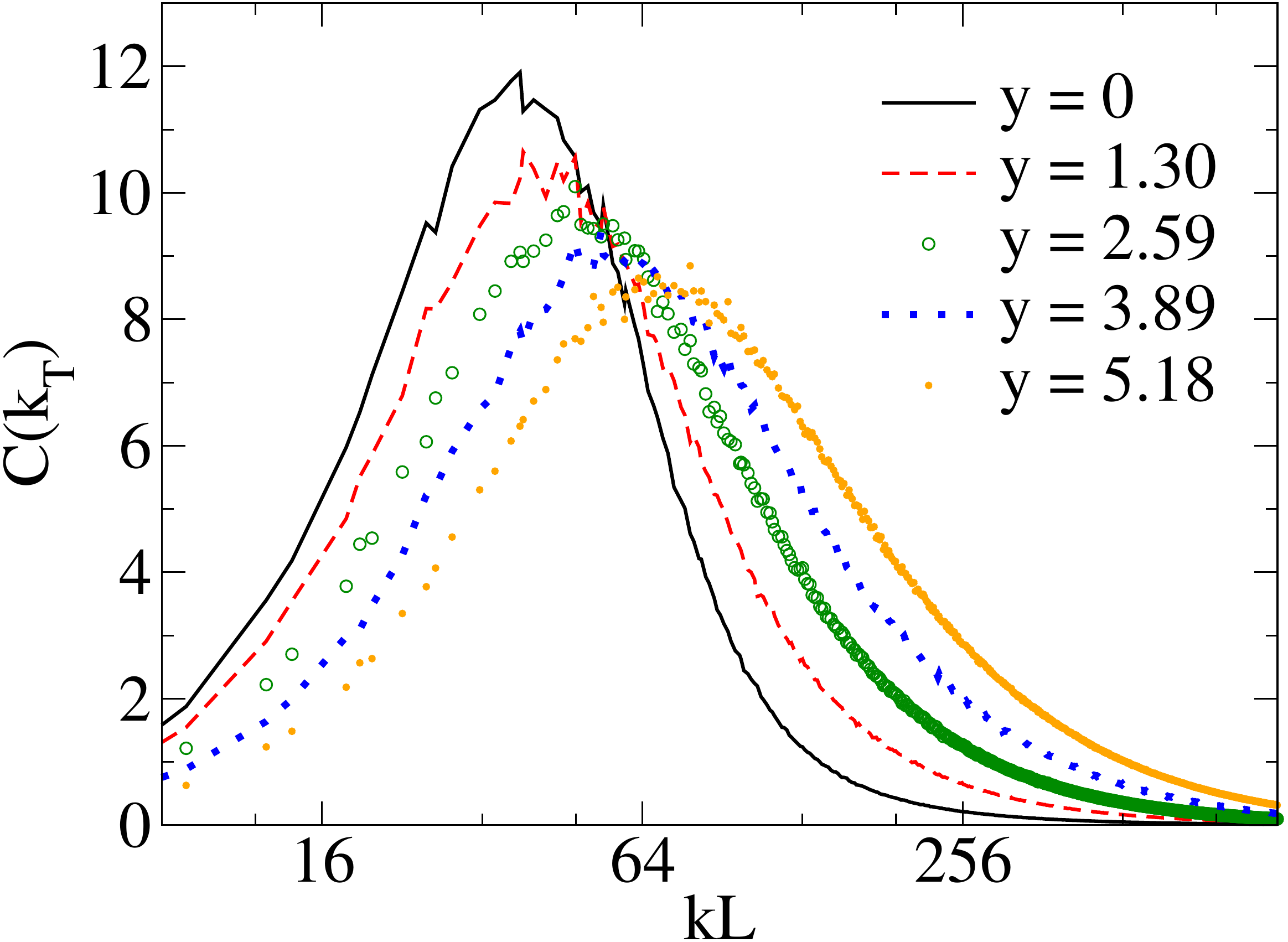}
\includegraphics[width=0.4\textwidth]{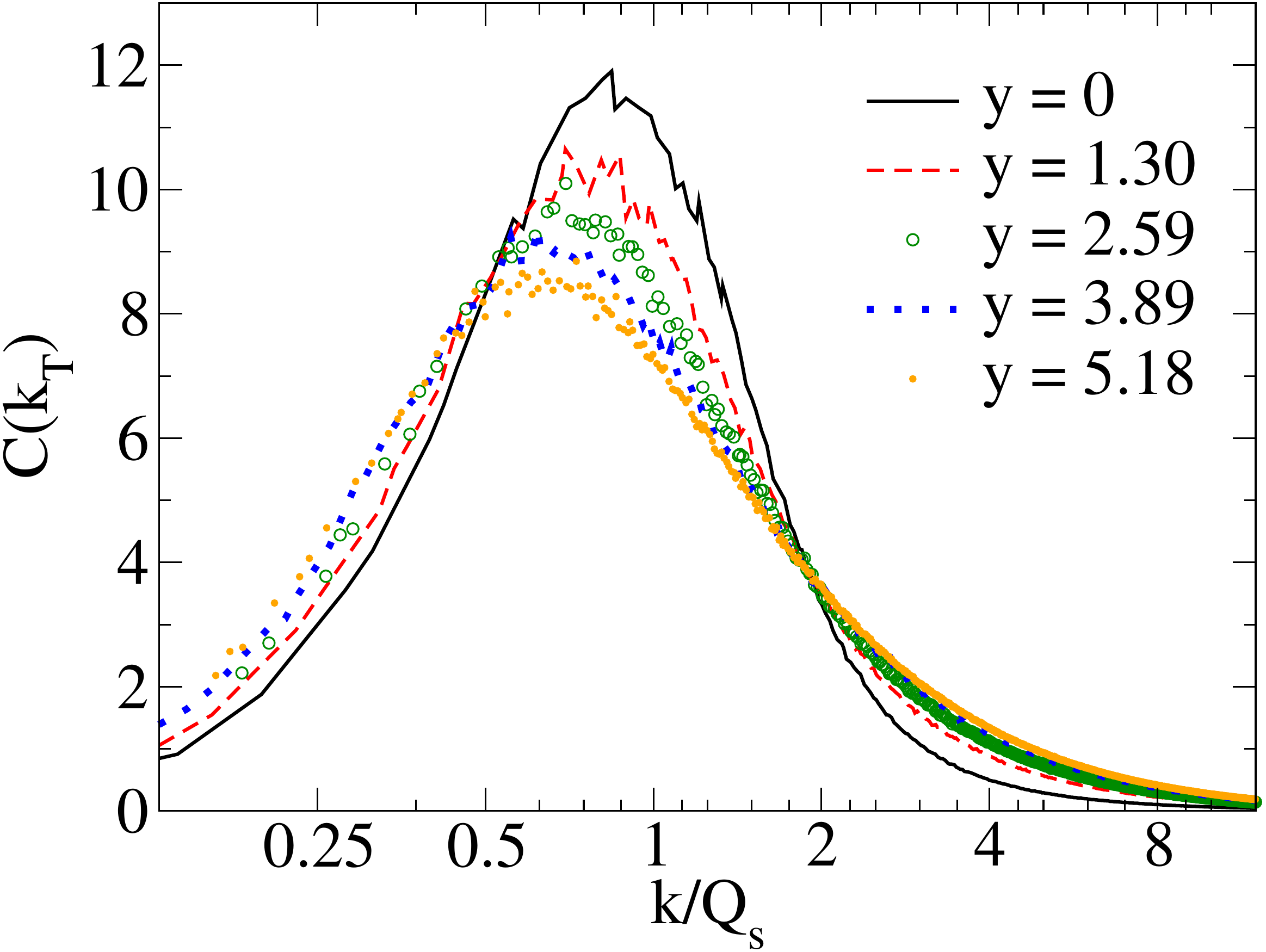}
\end{center}
\caption{
Wilson line correlator \nr{eq:defcorr} (dipole cross section) in momentum space; 
as s function of $\ktt L$ (above) and of the scaling variable $k/\qs$ (below).
} \label{fig:uuspect}
\end{figure}

The most elementary observable to monitor during the evolution
is the  Wilson line correlator
\begin{equation}\label{eq:defcorr}
C(r=|\xt-\yt|) = \frac{1}{\nc} \left\langle \tr U^\dag(\xt) U(\yt) \right\rangle.
\end{equation}
The related dipole cross section $2\int \ud^2 \bt (1-C(r))$ 
appears directly in the expression for 
the inclusive DIS cross section and is therefore of  interest in 
itself. The correlator 
varies between the values 1 at $r=0$ and 0 at large $r$, providing a 
natural way to define the saturation scale as the inverse of the correlation 
length of the Wilson lines as an intermediate scale between these two regimes.
We adopt the definition, as in \cite{Kowalski:2003hm},  of the fundamental representation 
$\qs$ by the  criterion 
\begin{equation}\label{eq:defqs}
C(r=\sqrt{2}/\qs)= e^{-1/2}.
\end{equation}
 Note that while 
for a Gaussian (``GBW''\cite{Golec-Biernat:1998js}) correlator
this definition is equivalent to the momentum space one used in Ref.~\cite{Lappi:2007ku},
they need not give exactly the same values in the general case.

In $\ktt$-factorized calculations of gluon production in pA collisions one needs 
the unintegrated gluon distribution of the dense target. This is obtained from the
the Fourier transform of \eq\nr{eq:defcorr} multiplied by $\ktt^2$
\begin{equation}\label{eq:defcorrk}
C(\kt) = \kt^2 \int \ud^2\rt e^{i \kt \cdot \rt} C(\rt).
\end{equation}
The typical behavior of the unintegrated distribution \nr{eq:defcorrk}
is to start at zero for small $\ktt$ and have  maximum around $\ktt \sim \qs$.

In the MV model the unintegrated gluon distribution behaves as $C(\ktt) \sim 1/\ktt^2$
for large $\ktt$,
which corresponds to the integrated
gluon distribution $xg(x,Q^2)$ behaving as $\sim \ln Q^2$. The main effects 
of JIMWLK/BK evolution are the increase of the characteristic scale $\qs$ with energy and
and making the functional form of the unintegrated distribution less steep, 
$\sim 1/\ktt^{2 \gamma}$. For fixed coupling the anomalous dimension 
is~\cite{Iancu:2002tr,*Mueller:2002zm} $\gamma \approx 0.63$
and for running coupling numerical solutions~\cite{Albacete:2004gw} of the BK 
equation give $\gamma \approx 0.85$. Our present calculation is done on a
linear (as opposed to logarithmic) lattice, and cannot go to very 
large momenta before lattice ultraviolet cutoff effects are felt in the spectrum.
Therefore one cannot hope to get a very good numerical
evaluation of the anomalous dimension. 
The change in the behavior of the unintegrated 
gluon distribution is, however, clearly observable.
The Wilson line correlators at different rapidities are shown in 
\fig\ref{fig:uupos} in coordinate space and in \fig\ref{fig:uuspect} in momentum 
space, both in lattice units and a functions of
the scaling variables $r\qs$ and $k/\qs$. Both the increase of $\qs$
and the development of a geometric scaling region are very well visible.

\begin{figure}[tb]
\includegraphics[width=0.4\textwidth]{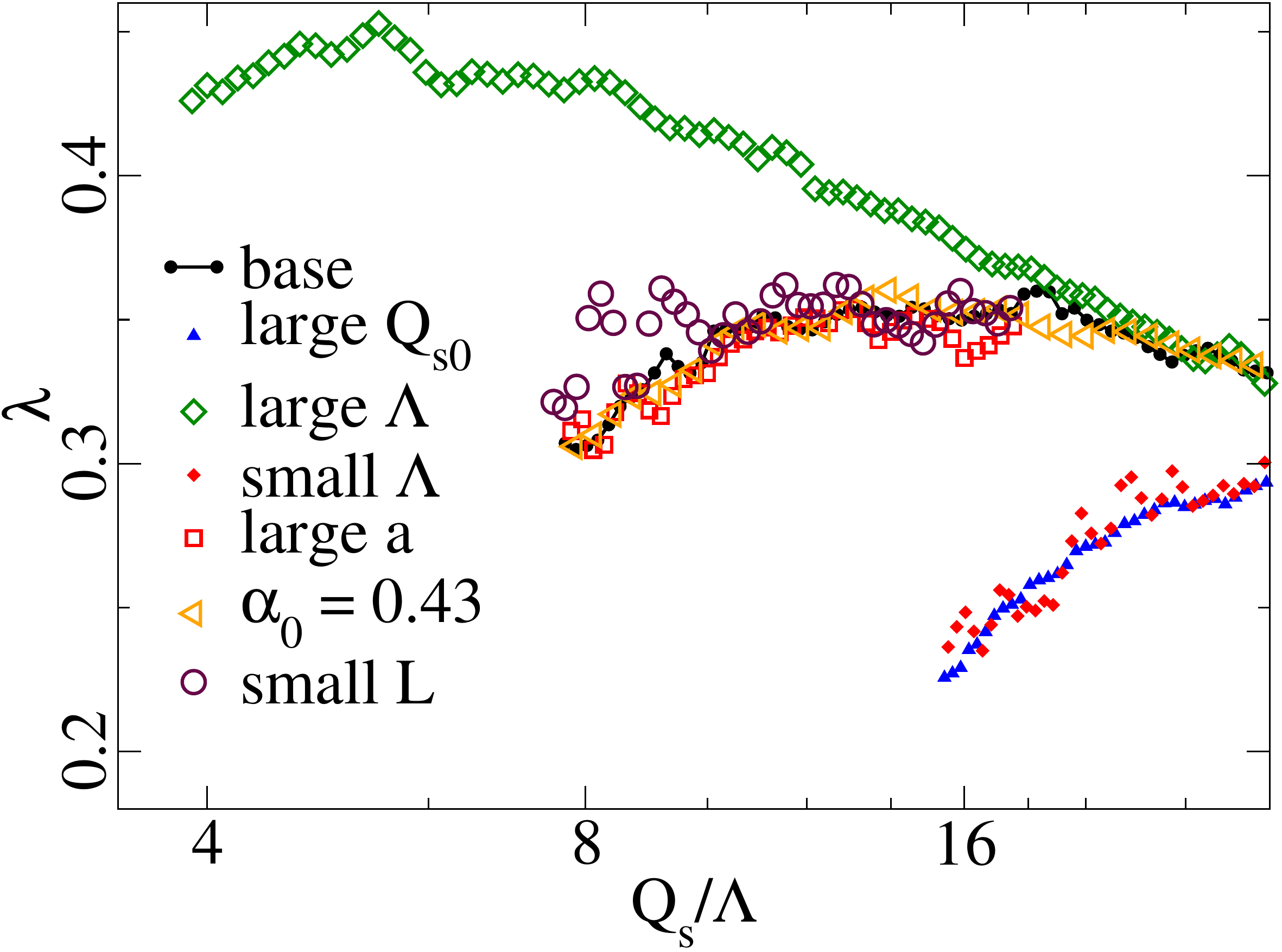}
\caption{
The evolution speed $\lambda=\ud \ln \qs^2(x)/\ud \ln 1/x$ as a function 
of the saturation scale $\qs$.
The parameter values corresponding to the labels are detailed in Table~\ref{tab:params}.
} \label{fig:lambdavsqs}
\end{figure}

\begin{figure}[tb]
\centerline{\includegraphics[width=0.45\textwidth]{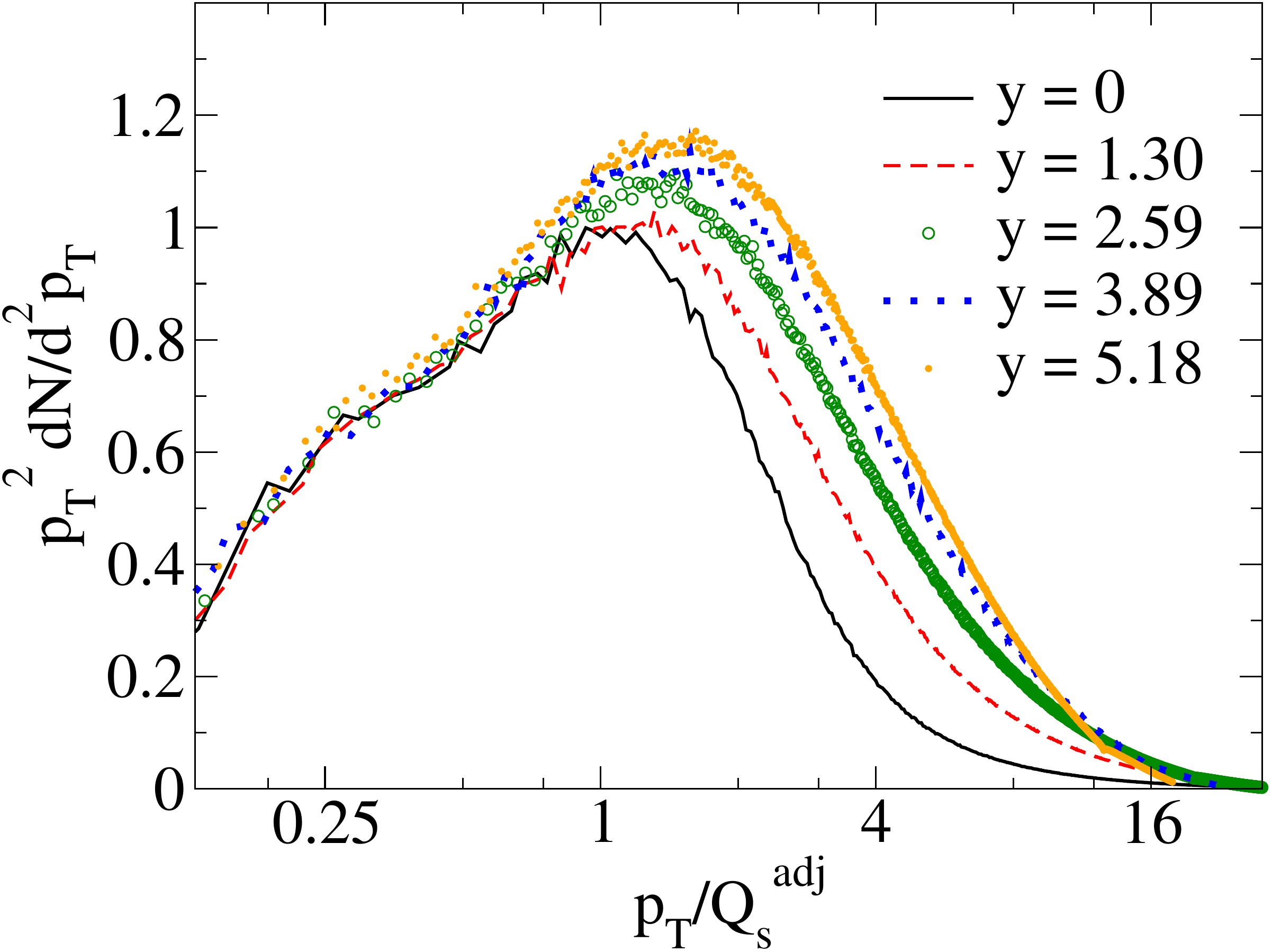}}
\caption{
Gluon spectrum at different energies, labeled by the rapidity interval of 
evolution starting from the MV initial condition  at $y=0$. The momentum is
scaled by the saturation scale $\qs$ corresponding to the rapidity in question.
} \label{fig:spectvsrap}
\end{figure}

\begin{table}
\begin{tabular}{|l||r|r|r|r|}
\hline
Configuration & $N_\perp$ & $\qsoadj L $  & $\mu_0 L$ & $\Lambda L$  \\
\hline
\hline
Base &   1024 & 68 & 15 & 6 \\
\hline
Large $\qso$ & 1024 & 136 &  15 & 6 \\
\hline
Large $\Lambda$ & 1024 & 68  &  30 & 12  \\
\hline
Small $\Lambda$ & 1024 & 68  &  7.5 & 3  \\
\hline
Large $a$ & 512 & 68 & 15 & 6 \\
\hline
Small $\alpha_0$ &   1024 & 68 & 30 & 6\\
\hline
Small $L$ & 512 & 34 & 7.5 & 3  \\
\hline
Small $L,a$ & 1024 & 34 & 7.5 & 3  \\
\hline
\end{tabular}
\caption{
The values of the numerical parameters used in our simulation sets. 
For $\mu_0/\Lambda=2.5$ the coupling is frozen at the value $\alpha_0=0.762$
and for $\mu_0/\Lambda=5$ at $\alpha_0=0.434$.
} \label{tab:params}
\end{table}

The values used in the different sets of numerical computations
in this paper are summarized in Table~\ref{tab:params}. 
The dependence of physical observables on the collision energy and rapidity 
depends most of all on the speed of evolution, conventionally parametrized as
\begin{equation}
\lambda = \frac{\ud \ln \qs^2(y)}{\ud y}.
\end{equation}
At fixed coupling $\lambda \sim \as$, and at running coupling the speed is expected to 
be proportional to $\as(\qs^2)$.
Thus the evolution speed is controlled by the relation of the initial saturation 
scale to the QCD scale controlling the running of the coupling, $\qsoadj/\Lambda$.
In a physically realistic case we would like to start the evolution at 
a scale corresponding to midrapidity at RHIC energies, i.e. 
$\qsoadj\approx 1.1\gev$~\cite{Lappi:2007ku},
corresponding to $\qsoadj/\Lambda\approx 11$, i.e. $\qso/\Lambda \approx 7.6$.
Assuming the transverse area to be $L^2=140\fm^2$ this leads to 
$\qsoadj L \approx 68$.
These correspond to the ``base'' set of values in Table~\ref{tab:params}
and used in \figs\ref{fig:uupos}, \ref{fig:uuspect} and~\ref{fig:spectvsrap}.
 To test the effect of the initial scale on 
the speed of evolution we also show results with two different values 
of $\qsoadj/\Lambda$. The change in $\qsoadj/\Lambda$ is obtained
by increasing $\qsoadj$ by a factor of 2 and keeping all other parameters fixed. 
Alternatively $\Lambda$ is decreased or increased  by a factor of 2, 
with everything else fixed.
The value at which the coupling freezes without altering the dynamics 
at higher momentum scales
can be altered by changing $\mu_0$. The sensitivity of the calculation 
to lattice effects has been tested 
by changing the lattice spacing $a$ (i.e. the lattice UV cutoff $\sim 1/a$)
and the physical volume $L=N_\perp a$ by a factor of two, with all the other 
dimensionful parameters fixed.

The initial conditions in the MV model
are constructed~\cite{Lappi:2007ku} 
as a product of $N_y=100$ infinitesimal Wilson lines, with the MV model
color charge density parameter $g^2\mu$ adjusted to provide the 
desired saturation scale.
The evolution speed for the different configurations is shown in 
\fig\ref{fig:lambdavsqs}, as a function of $\qs/\Lambda$. This confirms the expectation 
that starting with a lower $\qso/\Lambda$ results in a faster initial evolution. 
The phenomenologically preferred evolution speed $\lambda \lesssim 0.3$ is 
only reached with 
an initial saturation scale that is  higher than the above estimates corresponding 
to RHIC energy. Note, however, that this is affected by
the uncertainty concerning the correct value of 
$\Lambda$ in this running coupling scheme.

\section{Results for the gluon spectrum}
\label{sec:rescym}

We then take the Wilson line configurations from the JIMWLK evolution and, 
as discussed in Sec.~\ref{sec:method}, use them as initial conditions for 
the solution of the CYM equations of motion. The gluon spectrum resulting 
from the calculation is shown in \fig\ref{fig:spectvsrap}.
One can see that the gluon spectrum gets gradually harder as one moves from the 
initial condition into the geometric scaling regime. In the rapidity interval 
considered here  the total multiplicity and transverse energy of the 
gluons are still finite. This is to be contrasted with the
case of fixed coupling where, for an unintegrated gluon distribution 
behaving as $C(\ktt) \sim \ktt^{-2 \gamma}$ the produced gluon spectrum 
would behave as $\ud N/\ud^2 \pt \sim \ptt^{-4\gamma}.$ 
If the geometric scaling behavior continued to arbitrary large
$\ktt$ this would result in an ultraviolet divergent transverse energy for 
$\gamma < 0.75,$ including the fixed coupling value $\sim 0.63$.
 In practice the geometrical scaling region does not extend up to 
infinite $\ktt$, but a very hard gluon spectrum could be difficult to reconcile
with the observed transverse energy of the later stages of the quark gluon plasma.

\begin{figure}[tb]
\centerline{\includegraphics[width=0.4\textwidth]{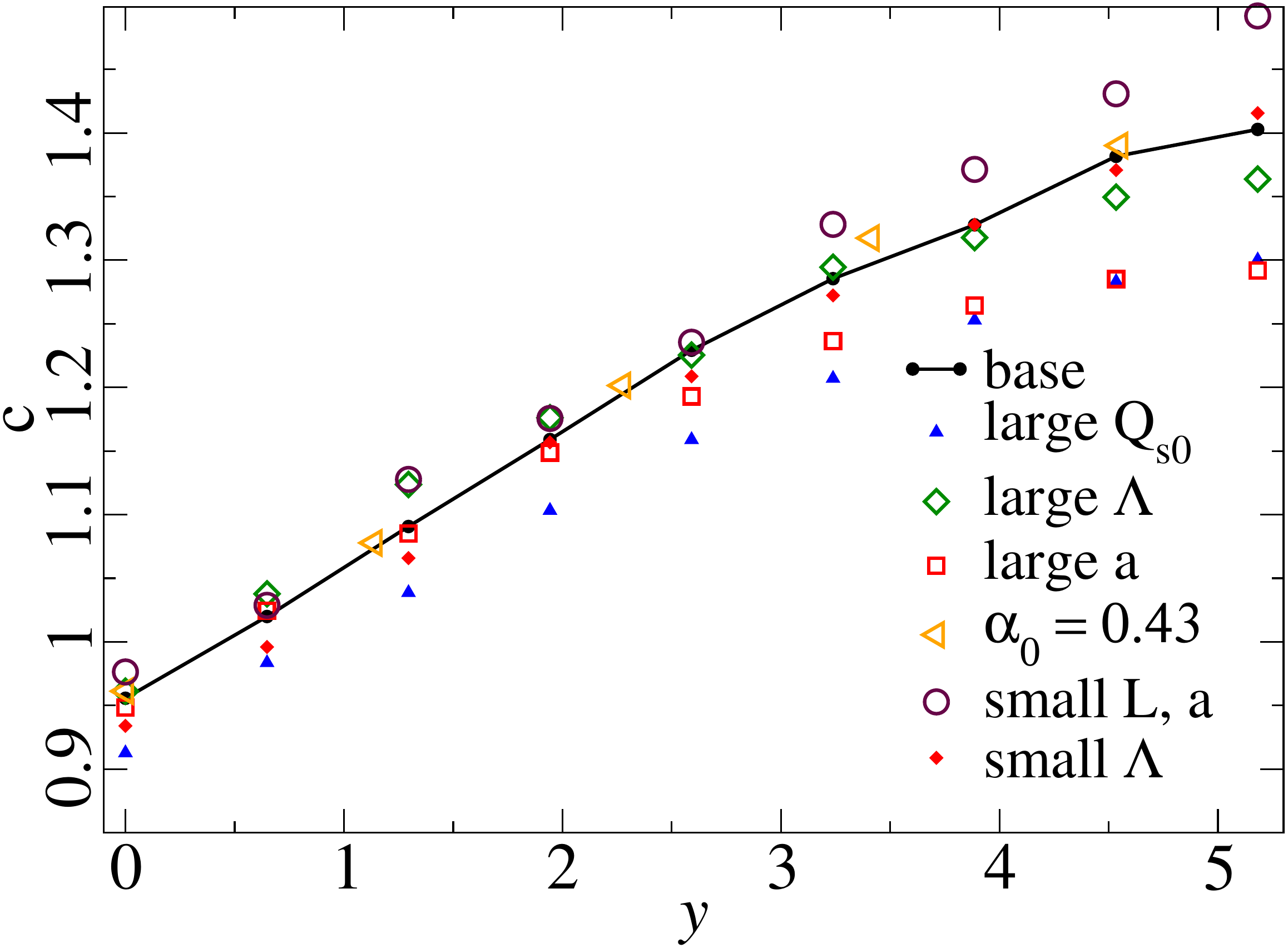}}
\caption{
Gluon liberation coefficient  as a function of collision energy, parametrized 
by the rapidity interval of evolution from the initial scale.
} \label{fig:cn}
\end{figure}

\begin{figure}[tb]
\centerline{\includegraphics[width=0.4\textwidth]{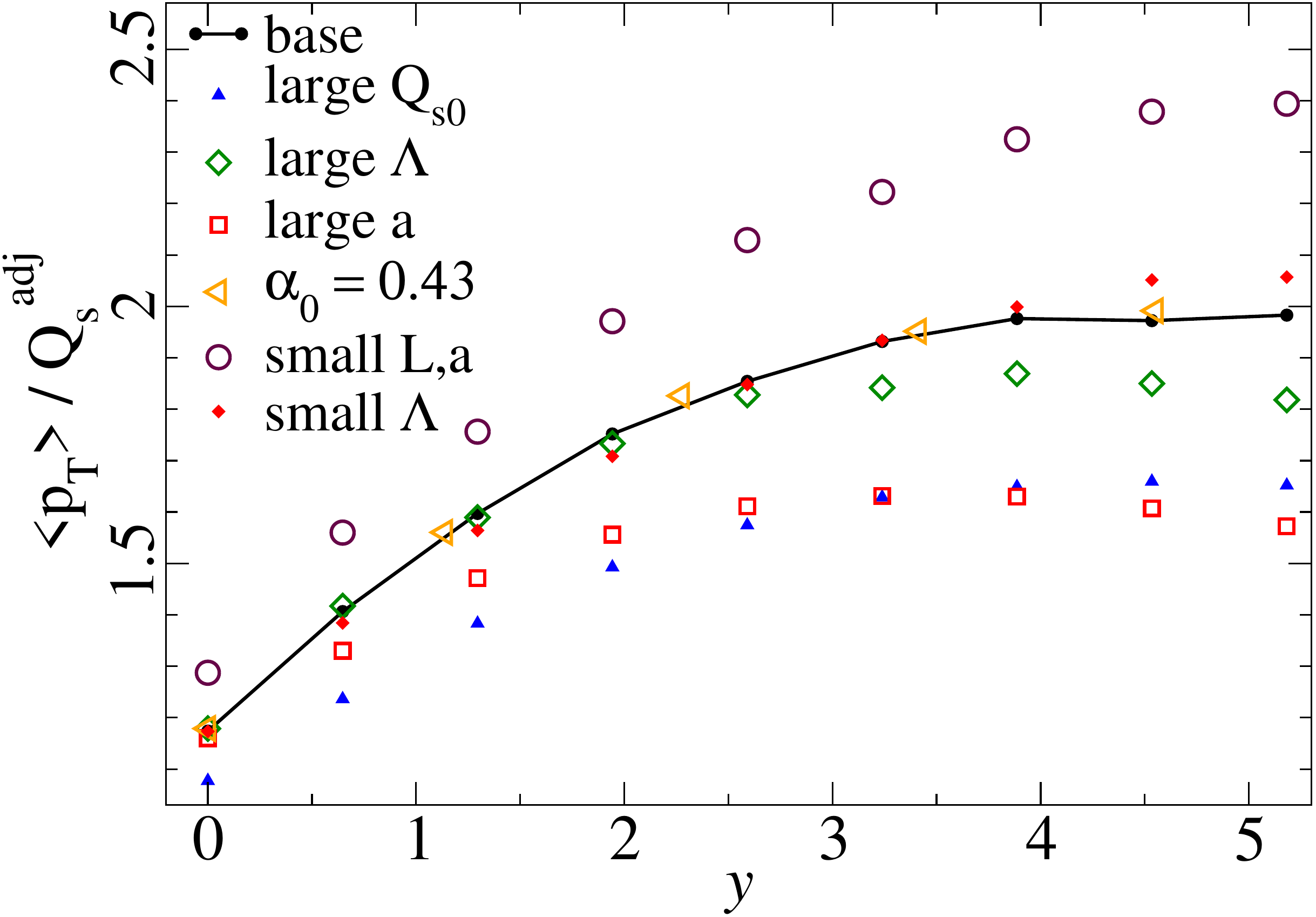}}
\caption{
Mean gluonic transverse momentum
as a function of collision energy..
} \label{fig:meanpt}
\end{figure}

The dominant transverse momentum scale of the produced gluon spectrum 
is expected to be the adjoint representation $\qsadj$.
It is convenient to parametrize the gluon spectrum by the 
dimensionless ``liberation 
coefficient''\cite{Mueller:1999fp,*Mueller:2002kw,Kovchegov:2000hz} 
$c$ proportional to the total gluon  multiplicity
\begin{equation} \label{eq:libc}
\frac{\ud N_{\mathrm{init. }g}}{\ud^2 \xt \ud y} = c \frac{\cf \qsadj^2}{2 \pi^2 \as},
\end{equation}
and the mean transverse momentum of the produced gluons
$\langle\ptt\rangle$. 
For a more thorough discussion on relating these values to 
the measured charged particle multiplicities we refer the reader to
e.g.~\cite{Lappi:2011gu}.

The values of $c$ and $\langle\ptt\rangle/\qsadj$  for
different amounts of evolution are plotted in \figs\ref{fig:cn} and~\ref{fig:meanpt}.
We see that, as expected, the harder $\ktt$-dependence in the initial wavefunctions
leads to a harder spectrum of produced gluons, as evidenced by the increase of 
$\langle\ptt\rangle/\qsadj$ from the initial condition. This increase does, however, 
seem to saturate after $y\approx 3$. This indicates that the functional form starts to 
settle towards a new scaling characteristic of JIMWLK evolution, where the 
anomalous dimension $\gamma < 1$ leads to a harder spectrum than in the MV model.
The mean transverse momentum of the gluons in the glasma, scaled by $\qs$,
increases from $\sim 1$ for the MV model initial condition  to around
$\sim 1.5$.

A somewhat less expected feature is the increase of the scaled multiplicity 
$c$ seen in \fig\ref{fig:cn}. This is seen also in the shape of the scaled spectrum 
in \fig\ref{fig:spectvsrap}, where new gluons are added for $\ptt \gtrsim 2\qsadj$,
but the shape for $\ptt \lesssim \qsadj$ changes very little. This 
observation could, as in Ref.~\cite{Blaizot:2010kh},
be interpreted as a difference between ``initial'' and ``final''
state interactions. At high transverse momentum $\ptt \gtrsim 2/\qsadj$
the produced gluon spectrum reflects the unintegrated gluon distributions
in the colliding projectiles. Thus $\ktt$-factorization works in this regime, and 
the spectrum gets harder because of the development of the geometrical 
scaling window.
The shape of the spectrum for $\ptt \lesssim \qsadj$, on the other hand, 
is a result of nonlinear interactions in the glasma stage, which render
the spectrum IR finite. These interactions are not captured in the
$\ktt$-factorized formalism.

Both $c$ and $\langle \ptt \rangle$
are smaller for the configurations where $\qs a$ is large (``small $a$'' and 
``large $\qs$''), which implies a strong dependence on the lattice
ultraviolet cutoff $1/a$. This is to be contrasted with \fig\ref{fig:lambdavsqs},
where the evolution speed $\lambda$ showed no significant dependence on 
the UV cutoff. 
Testing this by further decreasing $a$
with the same $L$ would become prohibitively expensive for this study.
We can, however, achieve smaller values of $\qs a$ by decreasing the size of the 
 system, i.e. $\qs L$. 
Making $\qso L$ smaller is eventually limited by the goal of staying in 
the strong field regime already for the initial rapidity.

\begin{figure}
\centerline{\includegraphics[width=0.4\textwidth]{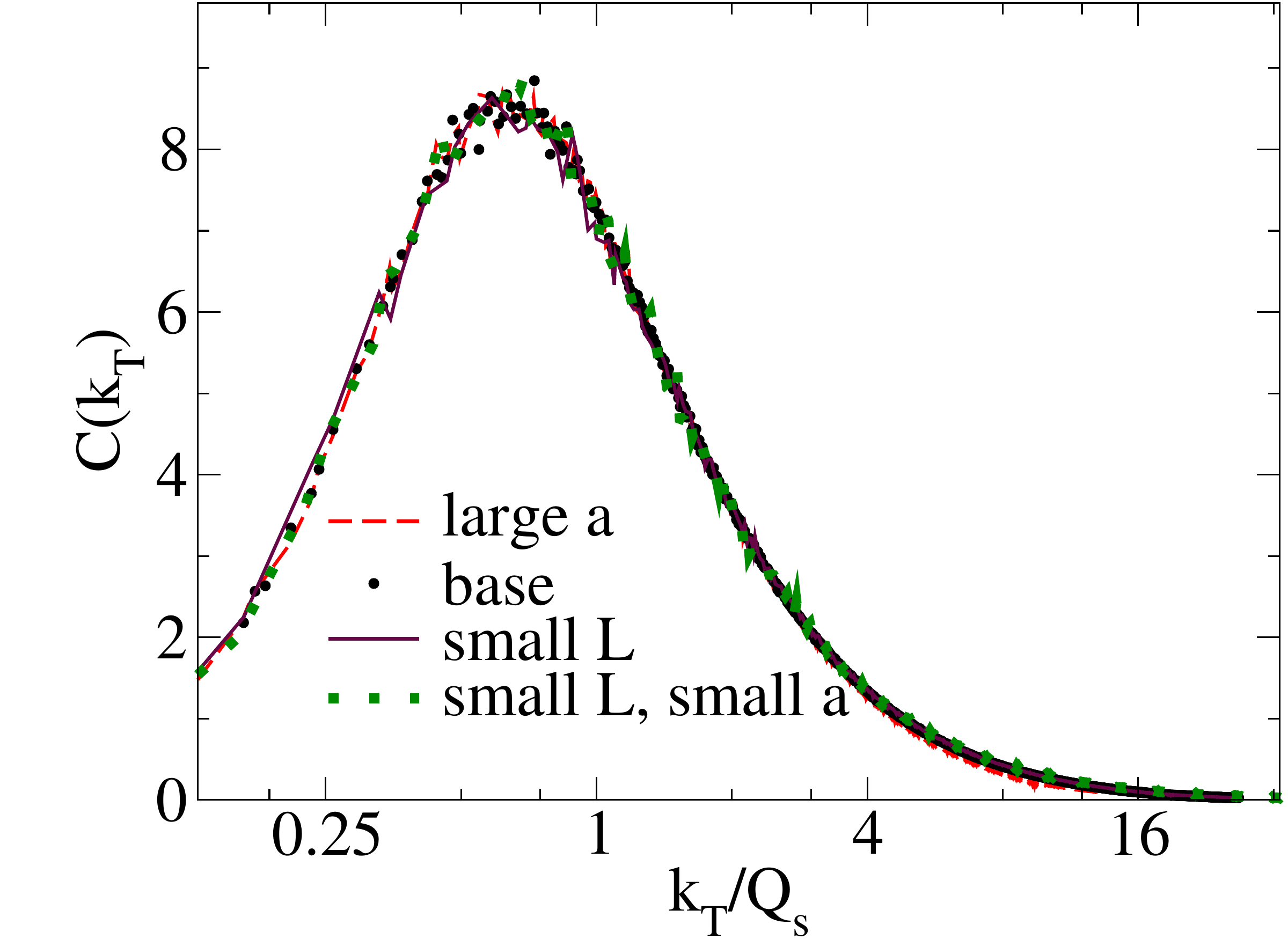}}
\caption{
The Wilson line correlator after $5.2$ units in rapidity. The
points and the dashed line differ only by the value of the
lattice spacing $a$ used. The dotted and solid line likewise differ only by the
value of $a$, and are obtained with $L$ half of the first ones.
For the parameter values used see Table~\ref{tab:params}.
} \label{fig:uuadep}
\end{figure}

\begin{figure}[!t]
\centerline{\includegraphics[width=0.4\textwidth]{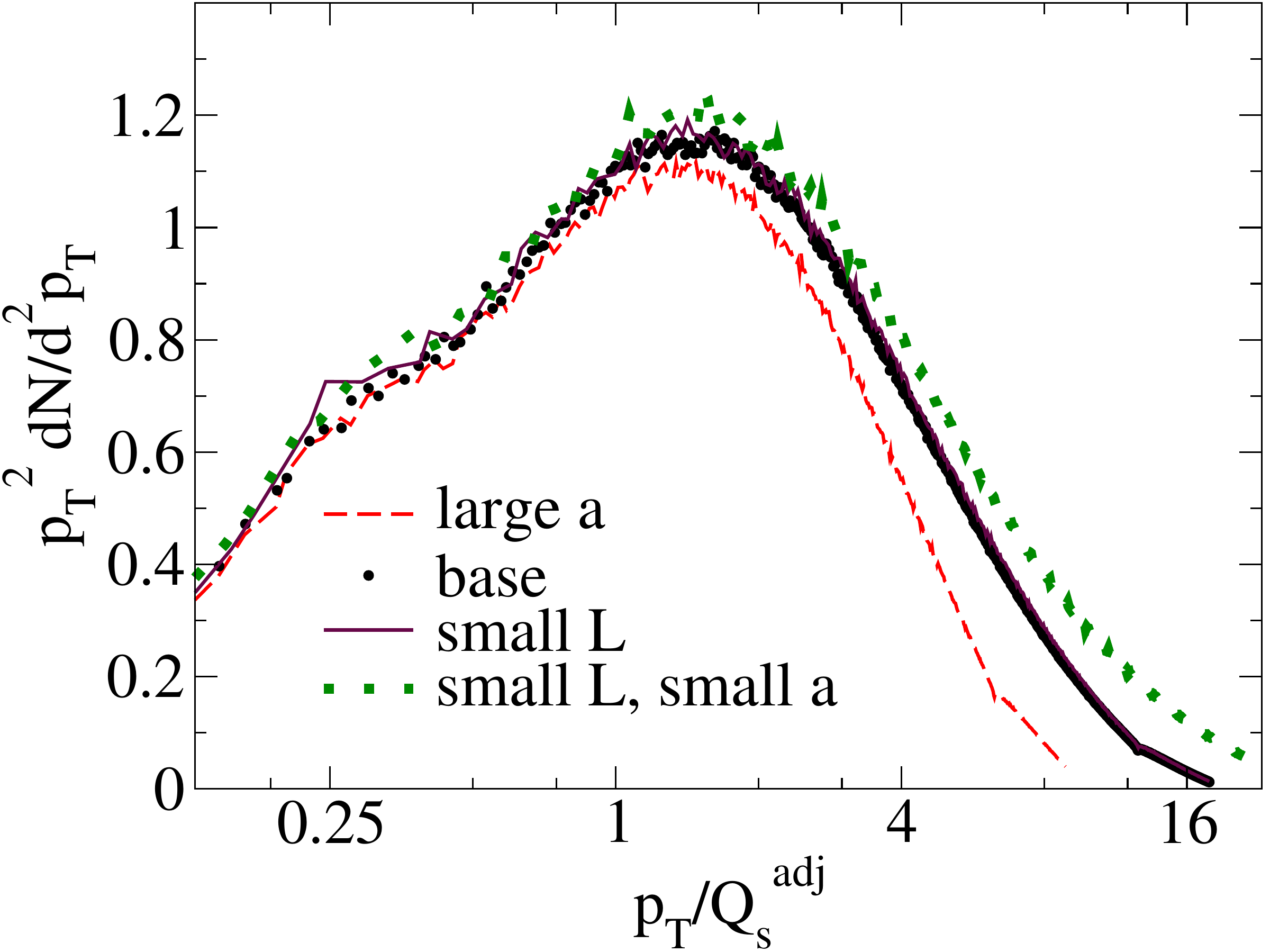}}
\caption{
Scaled gluon spectrum after $5.2$ units in rapidity. The labels are 
as in \fig\ref{fig:uuadep}.
} \label{fig:adep}
\end{figure}

The effect of the lattice UV cutoff is demonstrated
further in \figs\ref{fig:uuadep} and~\ref{fig:adep}. They show the 
unintegrated gluon distribution (\fig\ref{fig:uuadep}) and 
the produced gluon spectrum (\fig\ref{fig:adep}) after $5.2$ rapidity units
of evolution
for two different values of $\qso L$,
with two different values of $\qso a$ each, keeping the other parameters
of the evolution the same. One can see that with the new phase space opening up 
at smaller $a$ the unintegrated gluon distribution is mostly unaffected.
The produced gluon spectrum, on the other hand, increases in the high $\ptt$ tail. Thus 
the produced gluon spectrum is more sensitive to the lattice regularization 
than the JIMWLK evolution itself. One consequence of this is that the leveling 
off of the increase in $\langle\ptt\rangle/\qs$ at high rapidity, seen in 
\fig\ref{fig:meanpt}, could be influenced by lattice cutoff effects.

\section{Discussion}
\label{sec:disc}

We have, for the first time, used the solution of nonlinear
the high energy evolution equations in a calculation of gluon production 
in the initial stage of a heavy ion collision without resorting to
a $\ktt$-factorized approximation. This enables us to compute the gluon 
multiplicity and transverse energy density without ambiguities related
to the normalization and without additional infrared cutoffs.
It has been seen that the effect of JIMWLK evolution is to make the gluon
spectrum harder, leading to a growth of the total multiplicity that is
slightly faster than $\sim \qs^2$ and a gluon mean $\ptt$ that grows
faster than $\qs$. The resulting gluon spectrum in the glasma, shown in 
\fig\ref{fig:spectvsrap} is the main result of this paper.

The numerical calculation confirms that, as expected from 
studies of the BK equation,  introducing a 
running  coupling slows down the evolution
to a speed more consistent with experimental observations. 
At fixed coupling, numerical  JIMWLK evolution is known to be
very sensitive to the 
lattice ultraviolet cutoff. With running coupling the speed of evolution 
becomes essentially independent of the UV cutoff. However, when the solution is
used as an input for a calculation of the gluon spectrum of the initial glasma phase of 
heavy ion collisions, the lattice spacing dependence again becomes stronger.
A full systematic continuum extrapolation is, however, 
left for future work.
Effects of higher order corrections to JIMWLK/BK evolution  could significantly modify
the physics at $\ktt \gtrsim \qs.$ This would have a larger effect
on the spectrum of gluons in the glasma than expected just from the speed of the 
evolution.

The inclusion of NLO effects in our calculation has, by reasons of of numerical 
practicality, been limited to a ``daughter dipole'' prescription for the running coupling.
Incorporating more of the NLO corrections to JIMWLK/BK evolution into the
could have a much larger effect on  the gluon spectrum in a nucleus-nucleus collisions 
than on, say, the total DIS cross section. Another, separate but 
phenomenologically extremely topical issue that can be addressed 
in this same framework, are long range
rapidity correlations in the glasma~\cite{Dumitru:2008wn,Dusling:2009ni,Dumitru:2010iy}.
The calculation of the ``ridge'' correlation in the glasma proceeds in a 
similar fashion as the one performed in this paper, but is left for future work.

\section*{Acknowledgements}
Discussions with F. Gelis, K. Rummukainen, B. Schenke, R. Venugopalan and H. Weigert
are gratefully acknowledged.
This work  has been supported by the Academy of Finland, projects
126604 and 141555 and by computing resources from
CSC -- IT Center for Science in Espoo, Finland.

\rule{0pt}{5ex}

\rule{0pt}{5ex}

\bibliography{spires}
\bibliographystyle{JHEP-2modM}

\end{document}